\documentclass[prb,twocolumn,showpacs,preprintnumbers,amsmath,amssymb,floats,citeautoscript,nobalancelastpage]{revtex4}

\usepackage{graphicx}
\usepackage{dcolumn}
\usepackage{bm}
\usepackage{times}

\begin{document}

\title{Optical Study of the Electronic Structure and Correlation Effects in K$_{0.49}$RhO$_2$}

\author{R.~Okazaki$^{1}$}
\author{Y.~Nishina$^{1}$}
\author{Y.~Yasui$^{1}$}
\author{S.~Shibasaki$^{2}$}
\author{I.~Terasaki$^{1}$}

\affiliation{$^1$Department of Physics, Nagoya University, Nagoya 464-8602, Japan}
\affiliation{$^2$Department of Applied Physics, Waseda University, Tokyo 169-8555, Japan}

\date{\today}

\begin{abstract}

We study the optical properties of the layered rhodium oxide K$_{0.49}$RhO$_2$, which is isostructural to the thermoelectric material Na$_x$CoO$_2$.
The optical conductivity shows broad interband transition peaks as well as a low-energy Drude-like upturn, reminiscent of the optical spectra of Na$_x$CoO$_2$.   
We find that the peaks clearly shift to higher energies with respect to those of Na$_x$CoO$_2$, indicating a larger crystal-field splitting between $e_g$ and $t_{2g}$ bands in  K$_{0.49}$RhO$_2$.
The Drude weights suggest that the effective mass of K$_{0.49}$RhO$_2$ is almost two times smaller than that of Na$_x$CoO$_2$.
These differences in electronic structures and correlation effects between Na$_x$CoO$_2$ and K$_{0.49}$RhO$_2$ are discussed in terms of the difference between Co 3$d$ and Rh 4$d$ orbitals.

\end{abstract}



\maketitle

\section{introduction}

The layered cobalt oxide Na$_x$CoO$_2$ has attracted increasing research interests because of its highly unusual transport properties, $i.e.$ a large Seebeck coefficient with low electrical resistivity,\cite{Terasaki97} as well as a rich electronic phase diagram.\cite{Takada03,Motohashi03,Foo04}
Na$_x$CoO$_2$ consists of alternate stacking of Na and CoO$_2$ layers in which edge-sharing CoO$_6$ octahedra lead to a two-dimensional (2D) triangular lattice of Co ions,\cite{Terasaki97} in contrast to the 2D square lattice in the high-$T_c$ cuprates.
With increasing Na content $x$, the system varies from a Pauli-paramagnetic metal to a Curie-Weiss metal through a charge-ordered insulator around $x=0.5$.\cite{Foo04}
Furthermore, hydration induces a superconducting transition near $x=0.3$.\cite{Takada03}
The large Seebeck coefficient was observed at high Na doping region.\cite{Lee06}

To elucidate the origin of various intriguing phenomena in this material, the knowledge of the detailed electronic structure is crucially important.
The local density approximation (LDA) calculation predicted a large cylindrical hole Fermi surface and six small hole pockets,\cite{Singh00}
but the angle-resolved photoemission spectroscopy (ARPES) studies on Na$_x$CoO$_2$ found no small Fermi pockets.\cite{Hasan04,Yang04,Yang05,Qian06}
Furthermore, the measured bandwidths are significantly narrower than the LDA results, leading to a significance of correlation effects in this compound.
The LDA + Hubbard $U$ calculations indeed find a bandwidth renormalization and disappearance of the six small hole pockets,\cite{Zou04,Lee04,Zhang04,Zhang04b,Lee05}
but shows serious deviation from other experimental results as well.\cite{Johannes05} 
The optical conductivity spectra in Na$_x$CoO$_2$ well captured wide electronic structure including several interband transitions between Co $3d$ orbitals and the charge transfer from the low-lying O $2p$ orbitals as well as a low-energy anomalous Drude contribution.\cite{Lupi04,Wang04,Caimi04,Hwang05,Wu06} 
The measured conductivity peaks are able to be assigned to the LDA band structure, 
but the application of $U$ seriously  worsens the consistency between the LDA and the experimental spectra,\cite{Johannes05} 
puzzling a precise role of electron correlations for the electronic structure in this compound.

The isostructural $A_xB$O$_2$ ($A$ = alkali metal, $B$ = transition metal) provides valuable information to grasp a role of correlated Co $3d$ electrons in Na$_x$CoO$_2$.
In K$_{x}$CoO$_2$ system, various ground states are also expected with varying K content $x$,\cite{Sugiyama06,Sugiyama07}
although no superconductivity has been observed yet.\cite{Tang05}
Near $x=0.5$, K$_{x}$CoO$_2$ shows two distinct phase transitions at low temperatures,\cite{Watanabe06,Yokoi08} reminiscent of charge-ordered insulator Na$_{0.5}$CoO$_2$.\cite{Foo04}
The ARPES study also revealed common gapped features in the insulating phases of K and Na compounds near $x=0.5$.\cite{Qian06}
These experimental facts indicate similar electronic structures in these materials, as also inferred from the LDA calculations.\cite{Lee07}

In this paper, we present the optical study of the electronic structure in the $4d$-electron oxide K$_{0.49}$RhO$_2$ over a wide energy range.
This material shows a metallic conduction with relatively large Seebeck coefficient of 40 $\mu$V/K at room temperature,\cite{Shibasaki10} closely resembles metallic Na$_{x}$CoO$_2$.\cite{Terasaki97}
The $4d$-electron $A_x$RhO$_2$ system has also been investigated,\cite{Mendiboure87,Varela05,Park05,Krockenberger07} 
but its electronic state is poorly understood so far.
We find that whole optical conductivity spectrum of K$_{0.49}$RhO$_2$ is similar to that of Na$_x$CoO$_2$.
The locations of interband transition peaks, however, systematically shift to higher energies compared with those of Na$_x$CoO$_2$, indicating a larger crystal-field splitting between $e_g$ and $t_{2g}$ bands as well as wider bandwidth of $4d$ orbitals in an electronic structure of K$_{0.49}$RhO$_2$.
The spectrum weights also suggest a weak electron correlation in K$_{0.49}$RhO$_2$ due to a broad $4d$-orbital bandwidth: The effective mass of K$_{0.49}$RhO$_2$ is nearly half that of Na$_x$CoO$_2$.
In comparison with the optical spectra in Na$_x$CoO$_2$, the electronic structure and correlation effects in K$_{0.49}$RhO$_2$ are well described by the difference between Co 3$d$ and Rh 4$d$ orbitals.

\section{experimental}

Experiments have been performed on K$_{0.49}$RhO$_2$ single crystals with typical dimensions of $1\times1\times0.02$\,mm$^3$, grown by a self-flux method.\cite{Yubuta09} 
A mixture of K$_2$CO$_3$ and Rh$_2$O$_3$ with a molar ratio of 25:1 was kept at 1373 K for 1 hour, slowly cooled down to 1023 K with a rate of 2 K/hour.
As-grown samples were removed from the products by washing with distilled water.
The crystal is isostructural to $\gamma$-Na$_x$CoO$_2$ with the $P63/mmc$ symmetry, as illustrated in the inset of Fig.~1.

The $ab$-plane dc resistivity $\rho_{ab}$ was measured by a standard four-probe method.
The temperature dependence of $\rho_{ab}$ is depicted in Fig.~1. 
It has a typical Fermi-liquid behavior $\rho_{ab}(T) = \rho_0 + AT^2$ as shown in a previous report.\cite{Shibasaki10}
Near-normal-incidence $ab$-plane reflectivity spectra were measured at room temperature by using a Fourier-type interferometer Bomem DA3 equipped with an infrared microscope for energies between 0.1 eV and 2.5 eV.
To determine the absolute value of the reflectivity, we used a silver mirror for measuring the reference spectrum.
For a proper Kramers-Kronig (KK) analysis, the reflectivity measurements were extended up to 30 eV with the use of a Seya-Namioka-type grating monochromator for synchrotron radiation at BL-1B, UVSOR, Institute for Molecular Science.

\begin{figure}[t]
\includegraphics[width=7cm]{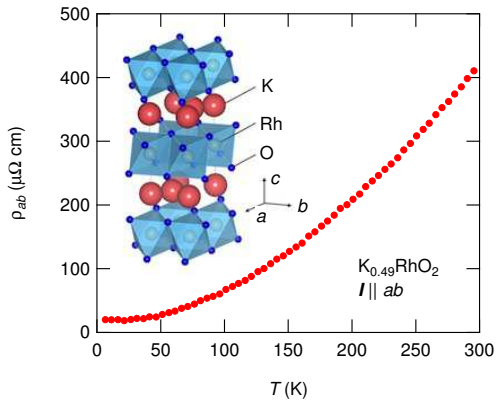}
\caption{(color online). Temperature dependence of the $ab$-plane resistivity $\rho_{ab}$ of K$_{0.49}$RhO$_2$ single crystal. Inset illustrates the crystal structure of K$_{x}$RhO$_2$. }
\end{figure}

\section{RESULTS AND DISCUSSION}

Figure\:2 shows the $ab$-plane reflectivity spectrum measured at room temperature.
Several broad peaks due to interband transitions are observed in a wide energy range, as also seen in reflectivity spectra of Na$_{x}$CoO$_2$.\cite{Lupi04,Wang04,Caimi04,Hwang05,Wu06} 
The metallic nature found in the resistivity is also confirmed by a low-energy Drude-like spectrum below the sharp edge around 1.2 eV, slightly higher than that of Na$_{x}$CoO$_2$ ($\simeq$ 0.7 eV).
We also show the high-energy reflectivity spectrum in the inset of Fig.~2.
Distinct absorption edge spectrum is observed near 9 eV, which derives from O $2p$ - Rh $4d$ $e_g$ charge-transfer transition as also observed in Na$_{0.7}$CoO$_2$.\cite{Caimi04}
The edge structure near 30 eV is probably a valence-electron plasma edge, generally observed in other transition-metal oxide materials such as high-$T_c$ cuprates.\cite{Uchida91}

The optical conductivity $\sigma_1$ is obtained from the KK analysis of the reflectivity.
In low energies, we extrapolated the reflectivity by using the Hagen-Rubens relation.
Standard extrapolation of $\omega^{-4}$ dependence was employed above 30 eV.
In Fig.~3(a), we show the optical conductivity spectrum below 8 eV.
The peak locations are well defined near 0.9 eV, 3.1 eV, and 5.5 eV as labeled by $\gamma'$, $\beta'$, and $\alpha'$, respectively.
We note that the peak shapes and positions are negligibly affected by above extrapolations.
For comparison, we also depict the room-temperature optical conductivity spectra of Na$_{x}$CoO$_2$ in Fig.~3(b), which were taken from previous reports.\cite{Wang04,Hwang05} 
These spectra depend on the Na content and show difference even in crystals with same composition,\cite{Wang04,Caimi04} but share the peak features in common.
In Na$_{x}$CoO$_2$, three conductivity peaks labelled by $\gamma$, $\beta$, and $\alpha$ are clearly recognized at 0.5 eV, 1.6 eV, and 3 eV, respectively.
In spite of little quantitative discrepancy between the LDA calculated spectra and experimental results, the measured conductivity peaks are well associated with the LDA band structure.\cite{Johannes05} 
Within the LDA, the O $2p$ bands lie well below the Fermi level, indicating a weak hybridization between Co $3d$ and O $2p$ orbitals.
The crystal-field splitting between lower $t_{2g}$ and upper $e_g$ bands in the octahedral environment is estimated to be about 2 eV.
Within this scheme, the $\gamma$ peak corresponds to the transition between $t_{2g}$ bands and the $\beta$ peak to $t_{2g}$-$e_g$ transition.
The $\alpha$ peak is responsible for the charge-transfer transition from occupied O $2p$ to unoccupied $e_g$ states, as schematically illustrated in Fig.~3(c).

\begin{figure}[t]
\includegraphics[width=7cm]{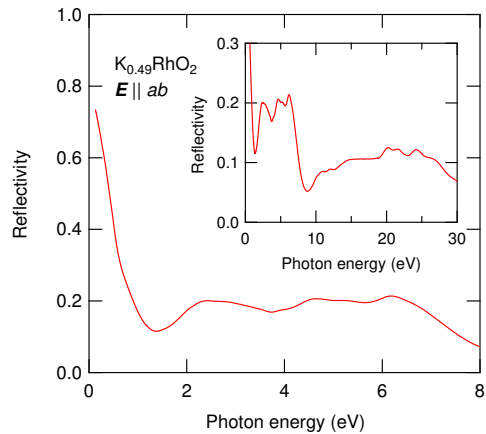}
\caption{(color online). Reflectivity spectrum of K$_{0.49}$RhO$_2$ single crystal measured at room temperature with polarization parallel to the $ab$ planes. Inset shows the reflectivity spectrum in high-energy region.}
\end{figure}

While the whole conductivity spectra among these Rh and Co oxides possess the similarities featured by three transition peaks and a low-energy Drude contribution, the peak positions in K$_{0.49}$RhO$_2$ are clearly shifted to higher energies relative to the spectra of  Na$_x$CoO$_2$.
It is also found that the peak widths are broader than those of Na$_x$CoO$_2$.
These difference are naively captured by a large crystal-field splitting and a wide bandwidth due to broader 4$d$ orbitals in K$_{0.49}$RhO$_2$.
Schematic energy diagram for K$_{0.49}$RhO$_2$ is depicted in Fig.~3(d).
A large crystal-field splitting pushes the $\alpha$ and $\beta$ peaks in Na$_x$CoO$_2$ to higher-energy $\alpha'$ and $\beta'$ peaks in K$_{0.49}$RhO$_2$, respectively.
The bandwidth of $t_{2g}$ complex (one $a_{1g}$ and two $e_g'$ bands) is expanded by broad orbitals of Rh $4d$ electrons, leading to a higher-energy shift of the $\gamma$ peak as well.

\begin{figure}[t]
\includegraphics[width=7cm]{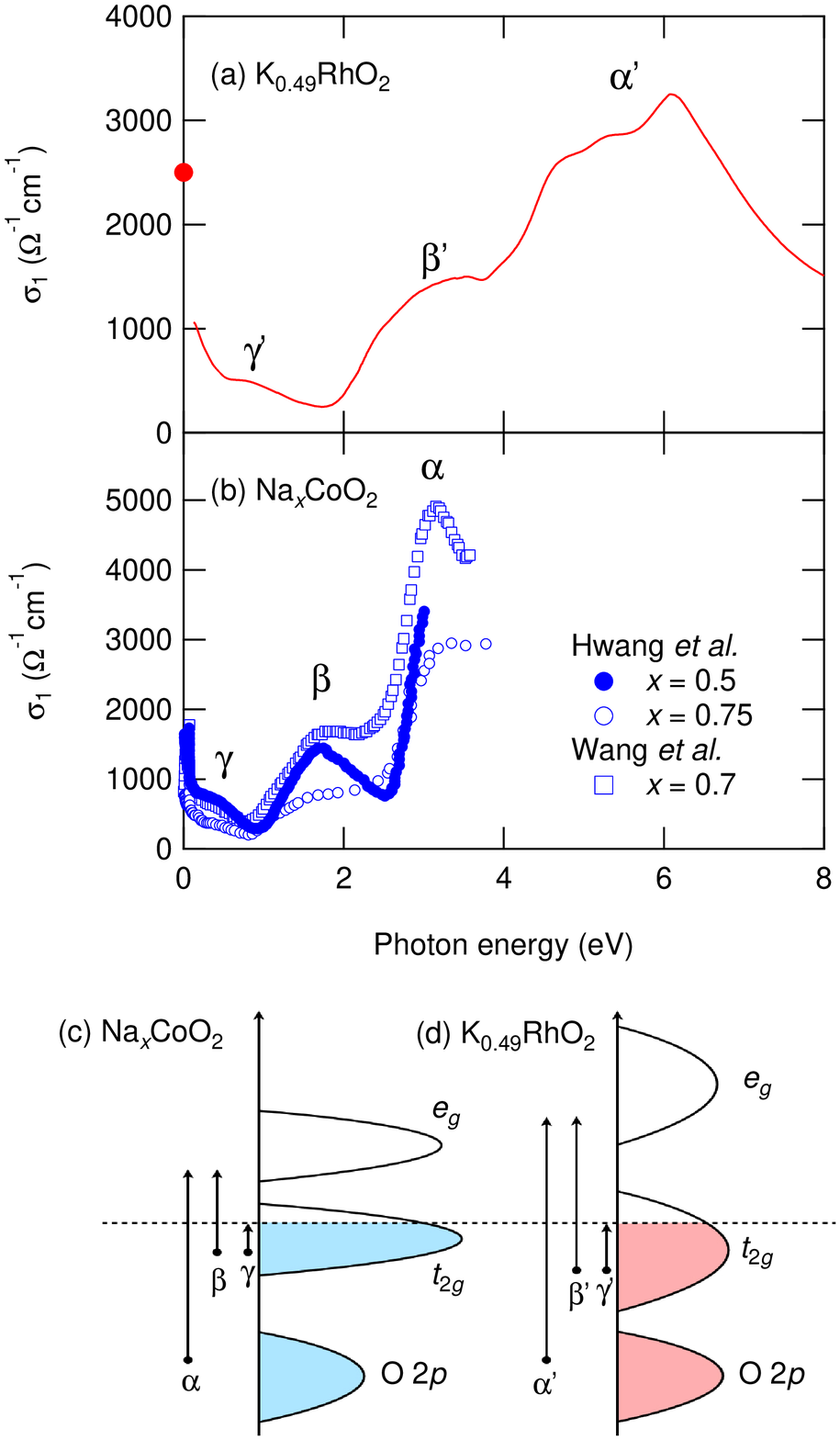}
\caption{(color online). Optical conductivity $\sigma_1(\omega)$ of (a) K$_{0.49}$RhO$_2$ obtained from the Kramers-Kronig transformation of the reflectivity spectrum and (b) Na$_x$CoO$_2$ taken from previous reports.\cite{Wang04,Hwang05}   The dc conductivity $\sigma_1(\omega \to 0)$  is plotted by the circle for K$_{0.49}$RhO$_2$. The peak locations in K$_{0.49}$RhO$_2$ (Na$_x$CoO$_2$) spectra are labelled by $\alpha'$ ($\alpha$), $\beta'$ ($\beta$), and $\gamma'$ ($\gamma$), which correspond to the O $2p$-$e_g$, $t_{2g}$-$e_g$ and $t_{2g}$-$t_{2g}$ transitions, respectively, as illustrated in schematic energy diagrams for (c) Na$_x$CoO$_2$ and (d) K$_{0.49}$RhO$_2$. }
\end{figure}

\begin{figure}[t]
\includegraphics[width=7cm]{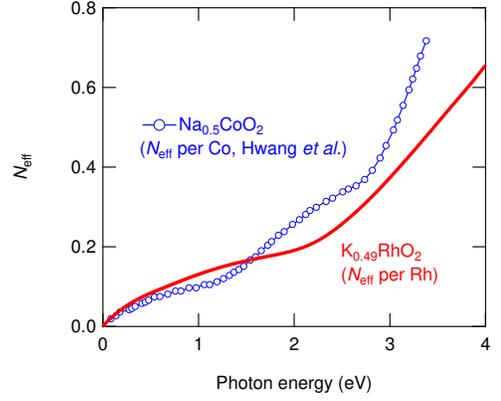}
\caption{(color online). Comparison of the effective carrier number in K$_{0.49}$RhO$_2$ and Na$_{0.5}$CoO$_2$ at room temperature. The data of Na$_{0.5}$CoO$_2$ is taken from Ref.\:\onlinecite{Hwang05}.}
\end{figure}

To evaluate the difference of electron correlation effects between these Co and Rh oxides, we have used the effective carrier numbers,  
\begin{equation}
N_{\rm eff}(\omega) = \frac{2m_0V}{\pi e^2}\int_0^{\omega}\sigma_1(\omega')d\omega',
\end{equation}
where $m_0$ is the free electron mass, $e$ is the charge of an electron, and $V$ is the volume occupied by one formula unit of K$_{0.49}$RhO$_2$.
Figure~4 displays $N_{\rm eff}(\omega)$ in K$_{0.49}$RhO$_2$.
Here we also plot the data for Na$_{0.5}$CoO$_2$, which is taken from Ref.\:\onlinecite{Hwang05}. 
The cut-off energy $\omega_c$ to separate the Drude weight from the $\beta$-band weight was adopted to be approximately 1.8 eV, where the conductivity spectrum exhibits distinct edge structure as shown in Fig.~3(a).
The Drude spectrum weight $N_{\rm eff}(\omega_c) \propto n/m^*$ ($n$ is the carrier concentration and $m^*$ is the effective mass)
is then given to be about 0.18 in K$_{0.49}$RhO$_2$,
nearly two times larger than that of Na$_{0.5}$CoO$_2$ [$N_{\rm eff}(\omega_c \sim 0.9$ eV) $\simeq$ 0.1].\cite{Hwang05}
The hole concentration $n$ in this isovalent family is determined by the alkali ion content $x$ as seen in the monotonic change of $N_{\rm eff}(\omega_c)$ with Na content $x$ in Na$_{x}$CoO$_2$,\cite{Hwang05}
which indicates almost same carrier concentrations among Na$_{0.5}$CoO$_2$ and K$_{0.49}$RhO$_2$. 
The effective mass in K$_{0.49}$RhO$_2$ is then nearly half that of Na$_{0.5}$CoO$_2$,
which is consistent with the $\gamma$-peak shift due to a $t_{2g}$ bandwidth spreading in K$_{0.49}$RhO$_2$ as mentioned above.

We here discuss the correlation effects in Na$_{x}$CoO$_2$.
While the LDA + $U$ calculation gives successful explanations for the measured electronic structure near the Fermi level, it cannot account for a wide electronic structure observed in the optical conductivity measurements.\cite{Johannes05} 
In contrast, the LDA result well reproduces the locations of conductivity peaks in Na$_{x}$CoO$_2$.
It is also suggested that a large effective mass due to strong quantum fluctuations corrects the slight difference between the LDA and experimental spectra.
Furthermore, quantum fluctuations may prevent the LDA-predicted magnetic ground state,\cite{Singh03}
 which is in sharply contrast to experimental facts.
In several correlated electron systems, the ground state varies from a magnetic order to a Fermi liquid through a quantum critical point (QCP) by increasing $W/U$, where $W$ is the bandwidth.
In this sense, Na$_{x}$CoO$_2$ with small $W$ locates near a QCP,  while K$_{0.49}$RhO$_2$ with large $W$ locates far away,
that is consistent with the substantial difference in the power of $\rho_{ab}(T)$ between these oxides.\cite{Shibasaki10,Li04}
The importance of quantum criticality in Na$_{x}$CoO$_2$ is also implied by magnetization measurements.\cite{Rivadulla06}

Besides these materials, several Co and Rh oxides have been intensively studied as potentially thermoelectric candidates.
In contrast to the present case, where the electronic structures are considerably different due to the difference between the Co $3d$ and Rh $4d$ orbitals, 
several compounds show little influence of the substitution for their electronic properties.\cite{Okada05,Hervieu06,Klein06}
In fact, the first-principle calculations revealed similar density-of-states spectra for NaCoO$_2$ and NaRhO$_2$ due to an accidental cancellation between a broader $4d$ orbital and a large ionic radius of Rh atom.\cite{Okada05}
On the other hand, in the energy range relevant to the optical conductivity measurements, one can recognize a slight difference of the bandwidths and crystal-field splitting between NaCoO$_2$ and NaRhO$_2$,\cite{Okada05} implying a considerable difference of optical spectra among K$_{0.49}$RhO$_2$ and Na$_{x}$CoO$_2$.
Further analysis for the measured optical conductivity of K$_{0.49}$RhO$_2$ based on a detailed band calculation should be required.

\section{CONCLUSION}

In summary, we have investigated the optical properties of the layered rhodate K$_{0.49}$RhO$_2$, isostructural to the thermoelectric material Na$_x$CoO$_2$.
We find the qualitative similarities in the optical conductivity spectra among these Rh and Co oxides, 
but the conductivity peaks in K$_{0.49}$RhO$_2$ evidently shift to higher energies with respect to those of Na$_x$CoO$_2$.
The peak shifts are naively explained by a large crystal-field splitting between $e_g$ and $t_{2g}$ bands and a broad $4d$-electron bandwidth in K$_{0.49}$RhO$_2$.
The comparison of Drude spectrum weights also indicates a significant difference of mass enhancement among these materials,
which is well consistent with the conductivity-peak shift due to a $4d$-bandwidth spreading in K$_{0.49}$RhO$_2$.

\section*{ACKNOWLEDGEMENTS}

We thank S. Kimura for collaboration and M. Hasumoto for technical assistance in UVSOR facility.
This work was supported by a Grant-in-Aid for Scientific Research from the Japan Society for the Promotion of Science.

\end{document}